\documentclass[12pt,letterpaper]{article}
\usepackage[margin=1in]{geometry}
\usepackage{color}
\usepackage{comment}
\usepackage{graphicx}
\usepackage{multicol,lipsum}
\usepackage[utf8]{inputenc}
\usepackage{aas_macros}
\usepackage{subfigure}
\usepackage{xcolor}
\usepackage{soul}
\usepackage{graphicx}
\usepackage{floatrow}
\usepackage{amsmath}
\usepackage[T1]{fontenc}
\usepackage{ amssymb }
\usepackage{xspace}
\usepackage{hyperref}
\usepackage{ragged2e}
\usepackage{sidecap}
\usepackage{floatrow}

\def\sgra{Sgr~A$^{\ast}$\xspace}        % Sgr A*
\def\m87{M87\xspace}                    % M87
\def\uv{$(u,v)$\xspace}                 % baseline coordinates (u,v) notation
\def\lsim{\mathrel{\raise.3ex\hbox{$<$\kern-.75em\lower1ex\hbox{$\sim$}}}}
\def\gsim{\mathrel{\raise.3ex\hbox{$>$\kern-.75em\lower1ex\hbox{$\sim$}}}}

\newenvironment{changemargin}[2]{%
\begin{list}{}{%
\setlength{\topsep}{0pt}%
\setlength{\leftmargin}{#1}%
\setlength{\rightmargin}{#2}%
\setlength{\listparindent}{\parindent}%
\setlength{\itemindent}{\parindent}%
\setlength{\parsep}{\parskip}%
}%
\item[]}{\end{list}}

\usepackage[maxbibnames=3,natbib=true]{biblatex}
\begin{document}

%%%% fancy coverpage, comment this out to get the original back %%%%%
\thispagestyle{empty}
\vspace{-0mm}
\begin{changemargin}{-.3cm}{-.3cm}
\begin{center}
    \vspace*{-3.2cm}\hspace*{-0.75cm}
    \makebox[\textwidth]{\includegraphics[width=\paperwidth,trim=0 1.5in 0 1.15in, clip]{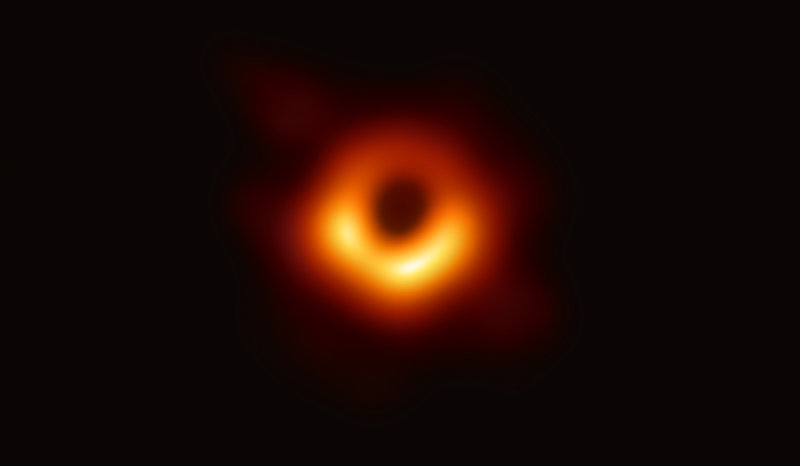}}
\end{center}
%%%%%% end of fancy coverpage markup %%%%%%\textsc{Name1}\\

\vspace{3mm}
\raggedright

\large
Astro2020 APC White Paper \linebreak
\vspace{-6mm}
% Title

\mbox{\textbf{Studying Black Holes on Horizon Scales with VLBI Ground Arrays}}\linebreak

\vspace{-2mm}
\small
\justify
%%%%% comment this out too vvvvv %%%%%%
% \vspace{2mm}
% \footnote[2]{Contact: dpesce@cfa.harvard.edu}
%%%%%% primary editors and organizers %%%%%%%
Lindy Blackburn$^{1, 2, \ast}$
% \affiliation{$^{1}$ Black Hole Initiative at Harvard University, 20 Garden Street, Cambridge, MA 02138, USA}
% \affiliation{$^{2}$ Center for Astrophysics $\vert$ Harvard \& Smithsonian, 60 Garden Street, Cambridge, MA 02138, USA}
Sheperd Doeleman$^{1, 2, \dagger}$,
% \affiliation{$^{1}$ Black Hole Initiative at Harvard University, 20 Garden Street, Cambridge, MA 02138, USA}
% \affiliation{$^{2}$ Center for Astrophysics $\vert$ Harvard \& Smithsonian, 60 Garden Street, Cambridge, MA 02138, USA}
%%%%%% major contributors to text, figures, MSRI source material %%%%%%%
Jason Dexter$^{12}$,
% \affiliation{$^{12}$ Max-Planck-Institut f\"ur Extraterrestrische Physik, Giessenbachstr. 1, D-85748 Garching, Germany}
Jos\'e L. G\'omez$^{16}$,
% \affiliation{$^{16}$ Instituto de Astrof\'{\i}sica de Andaluc\'{\i}a-CSIC, Glorieta de la Astronom\'{\i}a s/n, E-18008 Granada, Spain}
Michael D. Johnson$^{1, 2}$,
% \affiliation{$^{1}$ Black Hole Initiative at Harvard University, 20 Garden Street, Cambridge, MA 02138, USA}
% \affiliation{$^{2}$ Center for Astrophysics $\vert$ Harvard \& Smithsonian, 60 Garden Street, Cambridge, MA 02138, USA}
Daniel C. Palumbo$^{1, 2}$,
% \affiliation{$^{1}$ Black Hole Initiative at Harvard University, 20 Garden Street, Cambridge, MA 02138, USA}
% \affiliation{$^{2}$ Center for Astrophysics $\vert$ Harvard \& Smithsonian, 60 Garden Street, Cambridge, MA 02138, USA}
Jonathan Weintroub$^{1, 2}$,
% \affiliation{$^{1}$ Black Hole Initiative at Harvard University, 20 Garden Street, Cambridge, MA 02138, USA}
% \affiliation{$^{2}$ Center for Astrophysics $\vert$ Harvard \& Smithsonian, 60 Garden Street, Cambridge, MA 02138, USA}
%%%%%% other contributors %%%%%%%
% additions Nov 2019
Katherine L. Bouman$^{1, 2, 32}$,
Andrew A. Chael$^{1, 2, 33, 34}$,
% end of additions
Joseph R. Farah$^{1, 2, 21}$,
% \affiliation{$^{1}$ Black Hole Initiative at Harvard University, 20 Garden Street, Cambridge, MA 02138, USA}
% \affiliation{$^{2}$ Center for Astrophysics $\vert$ Harvard \& Smithsonian, 60 Garden Street, Cambridge, MA 02138, USA}
% \affiliation{$^{21}$
% University of Massachusetts Boston, 100 William T, Morrissey Blvd, Boston, MA 02125, USA}
Vincent Fish$^{4}$,
% \affiliation{$^{4}$ Massachusetts Institute of Technology, Haystack Observatory, 99 Millstone Road, Westford, MA 01886, USA}
Laurent Loinard$^{18, 19}$,
% \affiliation{$^{18}$ Instituto de Radioastronom\'{\i}a y Astrof\'{\i}sica, Universidad Nacional Aut\'onoma de M\'exico, Morelia 58089, M\'exico}
% \affiliation{$^{19}$ Instituto de Astronom\'{\i}a, Universidad Nacional Aut\'onoma de M\'exico, CdMx 04510, M\'exico}
Colin Lonsdale$^{4}$,
% \affiliation{$^{4}$ Massachusetts Institute of Technology, Haystack Observatory, 99 Millstone Road, Westford, MA 01886, USA}
Gopal Narayanan$^{28}$,
% $^{28}$ Department of Astronomy, University of Massachusetts, 01003, Amherst, MA, USA \\
Nimesh A. Patel$^{2}$,
% \affiliation{$^{2}$ Center for Astrophysics $\vert$ Harvard \& Smithsonian, 60 Garden Street, Cambridge, MA 02138, USA}
Dominic W. Pesce$^{1, 2}$,
% \affiliation{$^{1}$ Black Hole Initiative at Harvard University, 20 Garden Street, Cambridge, MA 02138, USA}
% \affiliation{$^{2}$ Center for Astrophysics $\vert$ Harvard \& Smithsonian, 60 Garden Street, Cambridge, MA 02138, USA}
Alexander Raymond$^{1, 2}$,
% \affiliation{$^{1}$ Black Hole Initiative at Harvard University, 20 Garden Street, Cambridge, MA 02138, USA}
% \affiliation{$^{2}$ Center for Astrophysics $\vert$ Harvard \& Smithsonian, 60 Garden Street, Cambridge, MA 02138, USA}
Remo Tilanus$^{17, 22, 23}$,
% \affiliation{$^{17}$ Department of Astrophysics, Institute for Mathematics, Astrophysics and Particle Physics (IMAPP), Radboud University, P.O. Box 9010, 6500 GL Nijmegen, The Netherlands}
% \affiliation{$^{22}$ Leiden Observatory---Allegro, Leiden University, P.O. Box 9513, 2300 RA Leiden, The Netherlands}
% \affiliation{$^{23}$ Netherlands Organisation for Scientific Research (NWO), Postbus 93138, 2509 AC Den Haag , The Netherlands}
Maciek Wielgus$^{1, 2}$,
% \affiliation{$^{1}$ Black Hole Initiative at Harvard University, 20 Garden Street, Cambridge, MA 02138, USA}
% \affiliation{$^{2}$ Center for Astrophysics $\vert$ Harvard \& Smithsonian, 60 Garden Street, Cambridge, MA 02138, USA}
%%%%%% comments, endorsers, EHT members %%%%%%
Kazunori Akiyama$^{1, 3, 4, 5}$,
% \affiliation{$^{1}$ Black Hole Initiative at Harvard University, 20 Garden Street, Cambridge, MA 02138, USA}
% \affiliation{$^{3}$ National Radio Astronomy Observatory, 520 Edgemont Road, Charlottesville, VA 22903, USA
% \affiliation{$^{4}$ Massachusetts Institute of Technology, Haystack Observatory, 99 Millstone Road, Westford, MA 01886, USA}
% \affiliation{$^{5}$ National Astronomical Observatory of Japan, 2-21-1 Osawa, Mitaka, Tokyo 181-8588, Japan}
Geoffrey Bower$^{6}$,
% \affiliation{$^{6}$ Institute of Astronomy and Astrophysics, Academia Sinica, 645 N. A'ohoku Place, Hilo, HI 96720, USA}
Avery Broderick$^{7, 8, 9}$,
% \affiliation{$^{7}$ Perimeter Institute for Theoretical Physics, 31 Caroline Street North, Waterloo, ON, N2L 2Y5, Canada}
% \affiliation{$^{8}$ Department of Physics and Astronomy, University of Waterloo, 200 University Avenue West, Waterloo, ON, N2L 3G1, Canada}
% \affiliation{$^{9}$ Waterloo Centre for Astrophysics, University of Waterloo, Waterloo, ON N2L 3G1 Canada}
Roger Deane$^{10, 11}$,
Christian M. Fromm$^{13}$,
% \affiliation{$^{13}$ Institut f\"ur Theoretische Physik, Goethe-Universit\"at Frankfurt, Max-von-Laue-Stra{\ss}e 1, D-60438 Frankfurt am Main, Germany}
Charles Gammie$^{14, 15}$,
% \affiliation{$^{14}$ Department of Physics, University of Illinois, 1110 West Green St, Urbana, IL 61801, USA}
% \affiliation{$^{15}$ Department of Astronomy, University of Illinois at Urbana-Champaign, 1002 West Green Street, Urbana, IL 61801, USA}
Roman Gold$^{13}$,
% \affiliation{$^{13}$ Institut f\"ur Theoretische Physik, Goethe-Universit\"at Frankfurt, Max-von-Laue-Stra{\ss}e 1, D-60438 Frankfurt am Main, Germany}
Michael Janssen$^{17}$,
% \affiliation{$^{17}$ Department of Astrophysics, Institute for Mathematics, Astrophysics and Particle Physics (IMAPP), Radboud University, P.O. Box 9010, 6500 GL Nijmegen, The Netherlands}
Tomohisa Kawashima$^{4}$,
% \affiliation{$^{4}$ Massachusetts Institute of Technology, Haystack Observatory, 99 Millstone Road, Westford, MA 01886, USA}
Thomas Krichbaum$^{29}$,
Daniel P. Marrone$^{20}$,
% \affiliation{$^{20}$ Steward Observatory and Department of Astronomy, University of Arizona, 933 N. Cherry Ave., Tucson, AZ 85721, USA}
Lynn D. Matthews$^{4}$,
% \affiliation{$^{4}$ Massachusetts Institute of Technology, Haystack Observatory, 99 Millstone Road, Westford, MA 01886, USA}
% \affiliation{$^{10}$ Department of Physics, University of Pretoria, Lynnwood Road, Hatfield, Pretoria 0083, South Africa}
% \affiliation{$^{11}$ Centre for Radio Astronomy Techniques and Technologies, Department of Physics and Electronics, Rhodes University, Grahamstown 6140, South Africa}
Yosuke Mizuno$^{13}$,
% \affiliation{$^{13}$ Institut f\"ur Theoretische Physik, Goethe-Universit\"at Frankfurt, Max-von-Laue-Stra{\ss}e 1, D-60438 Frankfurt am Main, Germany}
Luciano Rezzolla$^{13}$,
% \affiliation{$^{13}$ Institut f\"ur Theoretische Physik, Goethe-Universit\"at Frankfurt, Max-von-Laue-Stra{\ss}e 1, D-60438 Frankfurt am Main, Germany}
Freek Roelofs$^{17}$,
% \affiliation{$^{17}$ Department of Astrophysics, Institute for Mathematics, Astrophysics and Particle Physics (IMAPP), Radboud University, P.O. Box 9010, 6500 GL Nijmegen, The Netherlands}
\mbox{Eduardo Ros}$^{29}$,
Tuomas K. Savolainen$^{29, 30, 31}$,
Feng Yuan$^{24, 25, 26}$,
% \affiliation{$^{24}$ Shanghai Astronomical Observatory, Chinese Academy of Sciences, 80 Nandan Road, Shanghai 200030, People's Republic of China}
% \affiliation{$^{25}$ Key Laboratory for Research in Galaxies and Cosmology, Chinese Academy of Sciences, Shanghai 200030, People's Republic of China}
% \affiliation{$^{26}$ School of Astronomy and Space Sciences, University of Chinese Academy of Sciences, No. 19A Yuquan Road, Beijing 100049, People's Republic of China}
Guangyao Zhao$^{27}$
% \affiliation{$^{27}$ Korea Astronomy and Space Science Institute, Daedeok-daero 776, Yuseong-gu, Daejeon 34055, Republic of Korea}

\vspace{-3mm}

\begin{multicols}{2}
\noindent\scriptsize{\textit{$^{1}$ Black Hole Initiative at Harvard University, 20 Garden Street, Cambridge, MA 02138, USA \\
$^{2}$ Center for Astrophysics $\vert$ Harvard \& Smithsonian, 60 Garden Street, Cambridge, MA 02138, USA \\
$^{3}$ National Radio Astronomy Observatory, 520 Edgemont Road, Charlottesville, VA 22903, USA \\
$^{4}$ Massachusetts Institute of Technology, Haystack Observatory, 99 Millstone Road, Westford, MA 01886, USA \\
$^{5}$ National Astronomical Observatory of Japan, 2-21-1 Osawa, Mitaka, Tokyo 181-8588, Japan \\
$^{6}$ Institute of Astronomy and Astrophysics, Academia Sinica, 645 N. A'ohoku Place, Hilo, HI 96720, USA \\
$^{7}$ Perimeter Institute for Theoretical Physics, 31 Caroline Street North, Waterloo, ON, N2L 2Y5, Canada \\
$^{8}$ Department of Physics and Astronomy, Univ. of Waterloo, 200 University Avenue West, Waterloo, ON, N2L 3G1, Canada \\
$^{9}$ Waterloo Centre for Astrophysics, University of Waterloo, Waterloo, ON N2L 3G1 Canada \\
$^{10}$ Department of Physics, University of Pretoria, Lynnwood Road, Hatfield, Pretoria 0083, South Africa \\
$^{11}$ Centre for Radio Astronomy Techniques and Technologies, Department of Physics and Electronics, Rhodes University, Grahamstown 6140, South Africa\\
$^{12}$ Max-Planck-Institut f\"ur Extraterrestrische Physik, Giessenbachstr. 1, D-85748 Garching, Germany \\
$^{13}$ Inst. f\"ur Theoretische Physik, Goethe-Universit\"at Frankfurt, Max-von-Laue-Stra{\ss}e 1, D-60438 Frankfurt am Main, Germany \\
$^{14}$ Department of Physics, University of Illinois, 1110 West Green St, Urbana, IL 61801, USA \\
$^{15}$ Department of Astronomy, University of Illinois at Urbana-Champaign, 1002 West Green Street, Urbana, IL 61801, USA \\
$^{16}$ Instituto de Astrof\'{\i}sica de Andaluc\'{\i}a-CSIC, Glorieta de la Astronom\'{\i}a s/n, E-18008 Granada, Spain \\
$^{17}$ Department of Astrophysics, Institute for Mathematics, Astrophysics and Particle Physics (IMAPP), Radboud University, P.O. Box 9010, 6500 GL Nijmegen, The Netherlands \\ \\
$^{18}$ Instituto de Radioastronom\'{\i}a y Astrof\'{\i}sica, Universidad Nacional Aut\'onoma de M\'exico, Morelia 58089, M\'exico \\
$^{19}$ Instituto de Astronom\'{\i}a, Universidad Nacional Aut\'onoma de M\'exico, CdMx 04510, M\'exico \\
$^{20}$ Steward Observatory and Department of Astronomy, University of Arizona, 933 N. Cherry Ave., Tucson, AZ 85721, USA \\
$^{21}$ 
University of Massachusetts Boston, 100 William T, Morrissey Blvd, Boston, MA 02125, USA \\
$^{22}$ Leiden Observatory---Allegro, Leiden University, P.O. Box 9513, 2300 RA Leiden, The Netherlands \\
$^{23}$ Netherlands Organisation for Scientific Research (NWO), Postbus 93138, 2509 AC Den Haag , The Netherlands \\
$^{24}$ Shanghai Astronomical Observatory, Chinese Academy of Sciences, 80 Nandan Road, Shanghai 200030, PRC \\
$^{25}$ Key Laboratory for Research in Galaxies and Cosmology, Chinese Academy of Sciences, Shanghai 200030, PRC \\
$^{26}$ School of Astronomy and Space Sciences, Univ. of Chinese Academy of Sci., No. 19A Yuquan Road, Beijing 100049, PRC \\
$^{27}$ Korea Astronomy and Space Science Institute, Daedeok-daero 776, Yuseong-gu, Daejeon 34055, Republic of Korea \\
$^{28}$ Department of Astronomy, University of Massachusetts, 01003, Amherst, MA, USA \\
$^{29}$ Max-Planck-Institut f\"ur Radioastronomie, Auf dem H\"ugel 69, D-53121 Bonn, Germany \\
$^{30}$ Aalto University Department of Electronics and Nanoengineering, PL 15500, FI-00076 Aalto, Finland \\
$^{31}$ Aalto University Metsähovi Radio Observatory, Metsähovintie 114, FI-02540 Kylmälä, Finland \\
$^{32}$ California Institute of Technology, 1200 East California Boulevard, Pasadena, CA 91125, USA \\
$^{33}$ Princeton Center for Theoretical Science, Jadwin Hall, Princeton University, Princeton, NJ 08544, USA \\
$^{34}$ NASA Hubble Fellowship Program, Einstein Fellow \\
\line(1,0){250}\\
\textnormal{$^{\ast}$ \href{mailto:lblackburn@cfa.harvard.edu}{lblackburn@cfa.harvard.edu}, $^{\dagger}$\href{mailto:sdoeleman@cfa.harvard.edu}{sdoeleman@cfa.harvard.edu}} \\
% \noindent{}\textbf{Abstract:} The Event Horizon Telescope (EHT)
}}
\end{multicols}
\end{changemargin}
\newpage
\pagenumbering{arabic}

\section{Introduction}

This white paper outlines a process to design, architect, and implement a global array of 
radio dishes that will comprise a virtual Earth-sized telescope capable of making the first real-time movies of supermassive black holes (SMBH) and their emanating jets.  These movies will resolve the complex structure and dynamics at the event horizon, bringing into focus not just the persistent strong-field gravity features predicted by general relativity, but also details of active accretion and relativistic jet launching that drive galaxy evolution and may even affect large scale structures in the Universe.  SMBHs are the most massive and most compact objects predicted by Einstein’s theory of gravity. They are believed to energize the luminous centers of active galaxies, where they convert the gravitational potential energy of infalling matter to radiant power and jetted outflows of charged particles that can stretch to hundreds of thousands or even millions of light years. We propose to turn the extreme environment of their event horizons into laboratories where astronomers, physicists, and mathematicians can actively study the black hole boundary in real-time, and with a sensitivity and angular resolution that allow them to attack long-standing fundamental questions from new directions.

\begin{figure}[h]
    \centering
    \includegraphics[width=\textwidth]{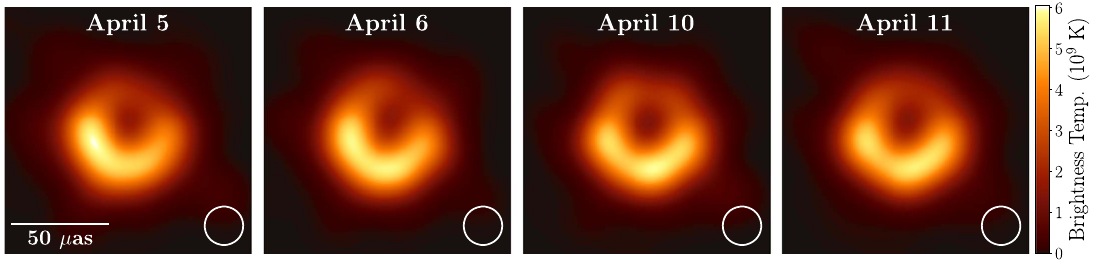}
    \caption{\footnotesize {1.3 mm wavelength images of \m87 for each of four days during which the source was observed in 2017 with the Event Horizon Telescope array.  All images are restored to an equivalent resolution with a beam of 20 $\mu$as.  These represent the highest angular resolution images ever made from the surface of the Earth, and show clearly the predicted photon orbit caused by extreme light bending in the presence of a 6.5 billion solar mass black hole.  The central dark region occurs because light rays interior to the photon ring spiral into the event horizon.  There is clear variation in the structure over the span of 5 days.}}
    \label{fig:m87_2017_results}
\end{figure}

While ambitious, this vision is grounded in recent remarkable results: on April 10th, 2019, after a decade of developmental efforts from a global collaboration of scientists \cite{paperii}, the Event Horizon Telescope (EHT) announced the first successful imaging of a black hole \cite{paperi}, \autoref{fig:m87_2017_results}. To accomplish this, the EHT used the technique of very long baseline interferometry (VLBI), in which radio dishes across the globe are synchronized by GPS timing and referenced to atomic clocks for stability, thereby synthesizing a single Earth-sized telescope.  Through development of cutting edge instrumentation, the EHT extended the VLBI technique to the 1.3 mm observing band, and by deploying these systems, created a virtual telescope with the highest angular resolution currently possible from the surface of the Earth. First experiments \citep{Doeleman_2008, Doeleman_2012} confirmed event horizon-scale structures in both \sgra and \m87.  Build-out of the full EHT enabled detailed imaging of the black hole ``shadow'' of \m87, which is formed by the lensed photon orbit of the 6.5 billion solar mass black hole at the galaxy's core.  Current EHT imaging capability is limited by the sparsity of VLBI baseline coverage, and targeted expansion of the EHT array 
by augmenting existing stations, as well as developing new sites, can greatly increase the scope of EHT core science over the next decade. We refer to the expanded array as the next-generation EHT, or ngEHT. A~separate white paper is dedicated to a complementary expansion of the EHT array by deployment of a radio telescope in orbit around the Earth.

\section{Key science goals and requirements}

The detection of the black hole shadow in \m87 \citep{paperi} has opened up the opportunity for repeated experimental studies of strong gravity and horizon scale accretion and jet launching with ngEHT. Future observations will measure the detailed shape and size of the black hole shadow and surrounding photon ring, allowing direct tests of the Kerr metric describing black holes in general relativity. An ngEHT will also address fundamental questions about the role of magnetic fields in the accretion and jet launching process as traced by the observed time-variable, polarized synchrotron radiation.

\begin{SCfigure}[5]
    \includegraphics[width=.65\textwidth,trim=0 1.5cm 0 0]{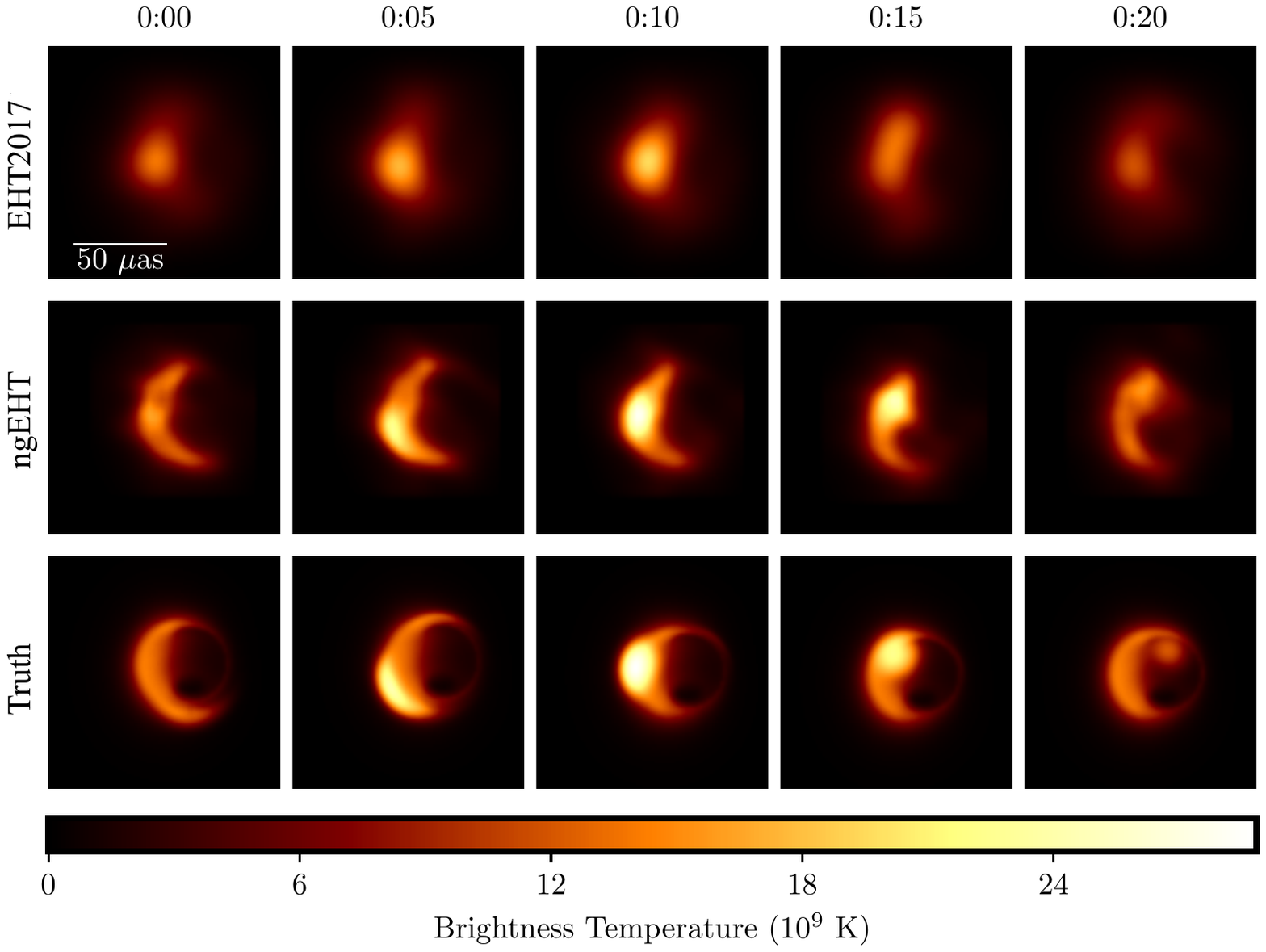}
    \caption{\footnotesize{Reconstructing movies of flares from \sgra with the EHT. Bottom row: Simulated images of a~``hot spot'' orbiting \sgra with a period of ${\sim}30$ minutes (Model B from \cite{Broderick_2006,Doeleman_2009}). 
    Upper rows: Corresponding reconstructions of the model with the EHT2017 and ngEHT arrays merging 230 and 345 GHz, demonstrating the potential to study the evolution of flares in \sgra on timescales of minutes \mbox{\cite{Johnson_2017,Bouman_2018}}. Reconstructions are performed with visibility amplitudes and closure phases, reflecting calibration similar to that of the EHT2017 data.}}
    \label{fig:sgra_movie}
\end{SCfigure}

\subsection{Testing General Relativity}

Measuring the shape and size of the shadow and surrounding lensed photon ring provides a null hypothesis test of General Relativity \citep{Johannsen_2010}. The mass and distance of \sgra are known to $\sim$1\% \citep{gravity_redshift_2018}, so the precision of GR tests for this source will be limited primarily by the fidelity of EHT data and the ability to extract the emission corresponding to the black hole circular photon orbit and interior shadow. Positional measurements of luminous matter (``hot spots'') orbiting near the event horizon, as shown in Fig.~\ref{fig:sgra_movie}, can be used to map the spacetime metric near the black hole and constrain the black hole spin. Combining one or more such EHT measurement of \sgra with other observations \citep[e.g.,][]{Gravity_2018} allows a test of the ``no hair'' theorem \citep{Hawking_1972,Broderick_2014,Psaltis_2015}.

Higher angular resolution allows more precise measurements of the shadow size and shape, while increased dynamic range improves image fidelity and allows us to extract the thin, bright photon ring feature from the more diffuse surrounding emission. Snapshot imaging of \sgra on timescales of minutes is required to track relativistic motions around the black hole. 

\subsection{The role of magnetic fields in black hole accretion}

\begin{SCfigure}[10][h]
    \includegraphics[width=0.68\textwidth]{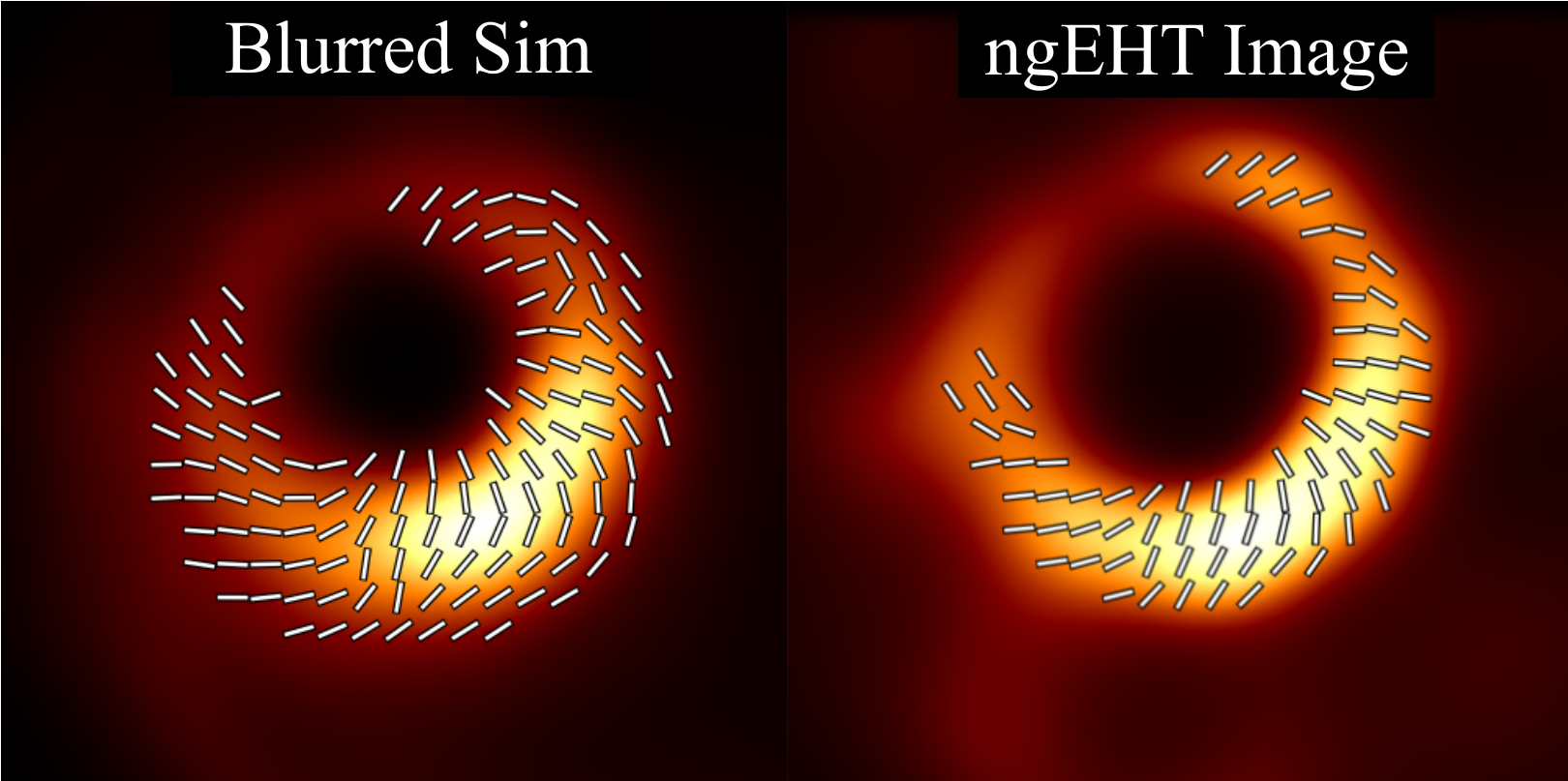}
    \caption{\footnotesize{Comparison of polarization map of a simulation of \m87 \cite{Chael_2019} blurred to half the nominal resolution of ngEHT (left), and a polarimetric reconstruction of synthetic ngEHT data generated by the simulation (right). ngEHT enables high fidelity polarimetric reconstructions, revealing the ordered, horizon-scale fields in this simulation of a ``magnetically-arrested'' disk.}}
    \label{fig:m87_pol}
\end{SCfigure}

Magnetic fields play an outsized role in accretion and jet formation. The magnetorotational instability \citep[MRI,][]{Balbus_1998} is thought to transport angular momentum and drive accretion onto the central black hole. Dynamically important magnetic fields can cause instabilities and flaring on horizon scales \citep{Tchekhovskoy_2011}. The  polarized synchrotron radiation observed by the ngEHT traces magnetic field geometry (Fig.~\ref{fig:m87_pol}), while its time variability encodes the dynamics of spiral waves driven by the MRI and magnetic flux eruption events associated with strong magnetic fields. Triggered multi-wavelength campaigns are needed to fully take advantage of ngEHT's capability to spatially resolve structures associated with the energetic, high energy flares from \sgra. The X-ray radiation in \sgra flares suggests that particles can be accelerated to high energy even around a quiescent black hole \citep{Dodds-Eden_2009}. Spatially resolving their radio counterparts will provide new constraints on the acceleration mechanism.

Snapshot polarimetric imaging with the future EHT can reveal the structure and dynamics of magnetic fields. Simultaneous polarimetric observations at 230 and 345 GHz will allow probing the magnetic field degree of ordering, orientation, and strength through Faraday rotation studies. Spectral index analyses will probe other plasma properties, such as the electron density and temperature.

\subsection{Jet formation}

The processes that govern the formation, acceleration and collimation of powerful relativistic jets in active galactic nuclei (AGN) and X-ray binaries are a half-century-long mystery in black hole physics. The leading scenarios rely on magnetic fields to extract rotational energy, either from orbiting material \citep{Blandford:1982vy} or from the black hole itself \citep{Blandford_1977}. Magnetic fields downstream further collimate and accelerate the jet to relativistic speeds. EHT observations of \m87 provide a unique opportunity to study jet launching, collimation, and acceleration at the base in the immediate vicinity of the black hole.

\begin{figure}[h]
    \centering
    \includegraphics[width=\textwidth]{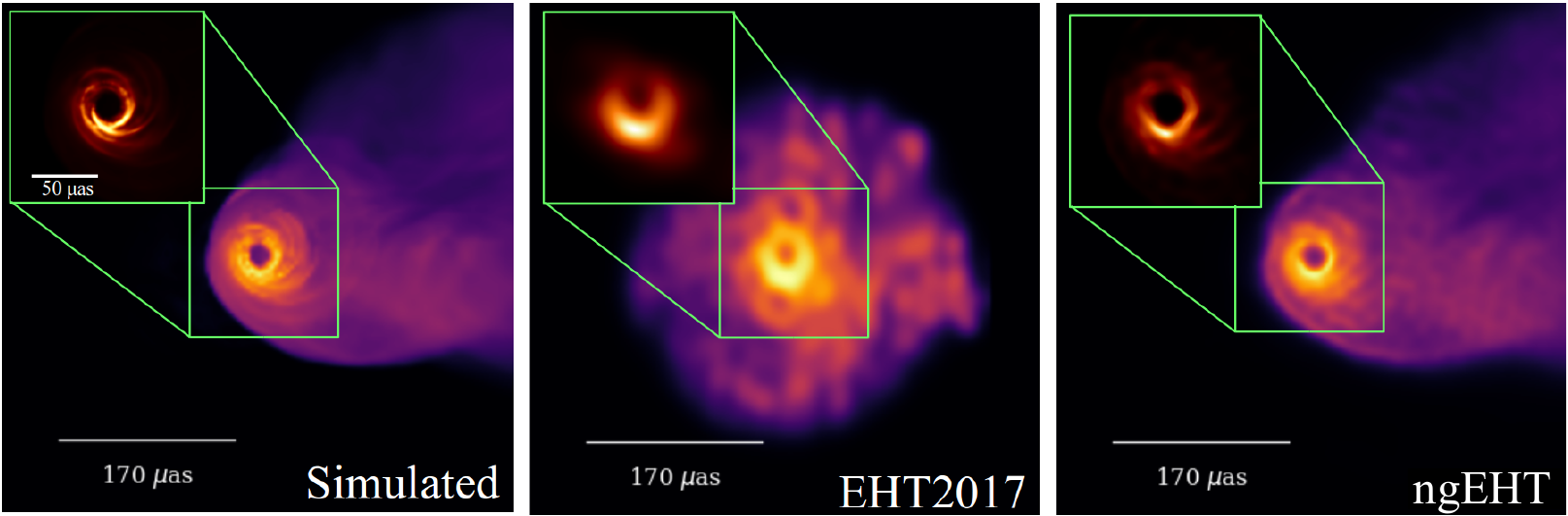}
    \caption{GRMHD snapshot from a simulation of \m87 (Chael et al. 2019). Main panel is log scale; inset is linear scale. Middle: Reconstruction using EHT2017, revealing the circular $\sim$ 40 $\mu$as ring surrounding the black hole shadow but no jet. Right: Reconstruction using ngEHT, including both 230 and 345 GHz, revealing both the black hole and its jet.
    }
    \label{fig:ngEHT_jet}
\end{figure}

Figure \ref{fig:ngEHT_jet} shows reconstructed 3D GRMHD simulations of the jet launching region in \m87 with current and sample ngEHT arrays. The addition of short baselines anchored to existing large apertures, combined with observations at progressively higher frequencies will improve both the imaging dynamic range and angular resolution to study formation, collimation and acceleration of relativistic jets, not only in M87 but also in other nearby AGN \citep{Boccardi_2017}. This opens new possibilities for linking jet power to black hole spin, accretion rate, and disk magnetization through direct comparison of observation and simulations on scales down to the event horizon. Triggered observations and centralized data processing will increase the cadence of the observations, allowing the study of time variable jet ejections first near the black hole, and subsequently as they emerge from the AGN core.

\subsection{Objectives and requirements}

The quality of ngEHT images/movies and their corresponding traction on key science questions depend on the baseline coverage of the array as well as overall sensitivity, observing frequency, bandwidth, and observing/scheduling constraints. Additional improvements in imaging and analysis algorithms will further drive design requirements and trade-offs in defining the ngEHT instrument systems and array architecture.  We advocate a formal system engineering approach, in which key science questions are used to define and select across technical elements for the array. In \autoref{tab:stm} we outline an abbreviated science traceability matrix for the ngEHT.  The full array design must be explored using system engineering driven simulation and science optimization process resulting in an expanded STM to define a phase of ngEHT design, followed by a phase of implementation, both timed to deliver a functional ngEHT array by the end of the coming decade.

\begin{table}
\begin{center}
\includegraphics[width=0.95\textwidth]{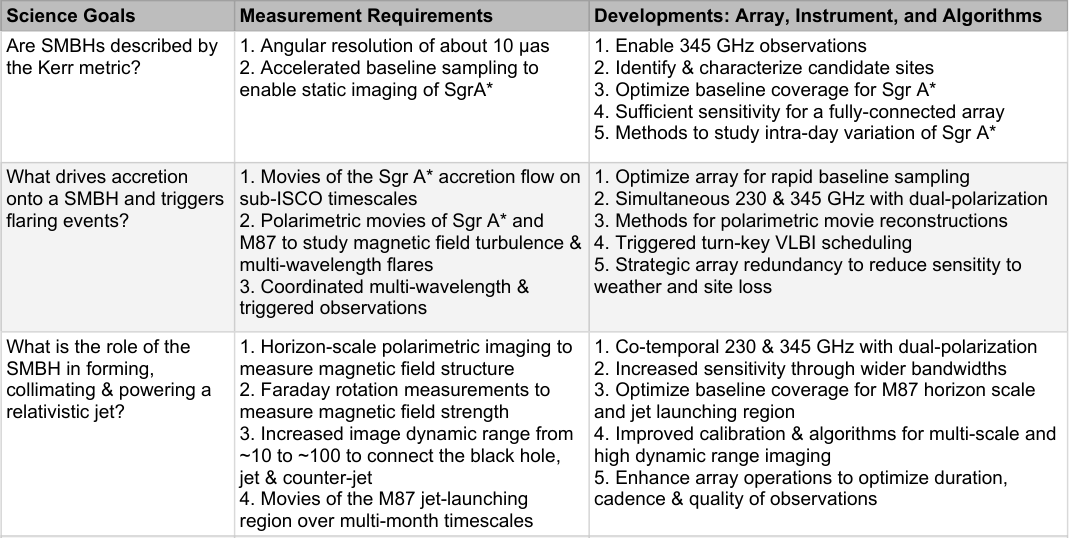}
\caption{A short form science traceability matrix (STM). The STM links the key science questions in the first column with the top level requirements for astronomical measurements in the second, and these drive the specifications for detailed design, of the array configuration, instrument developments, and software post processing algorithms in the third column.  System engineering will expand this STM in the early phases of an upgrade.}
\label{tab:stm}
\end{center}
\end{table}

\section{Technical elements}

\begin{SCfigure*}[50]
\includegraphics[width=0.5\textwidth,trim=0.5cm 1cm 0 0]{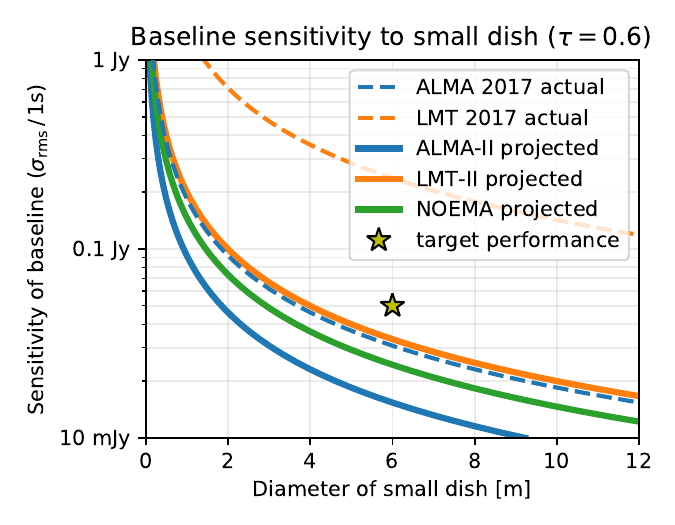}
\caption{\footnotesize {Key anchor stations in the ngEHT with sufficient sensitivity to connect small dishes to the entire array on $\sim$few second atmospheric timescales. A star marks a low level of correlated flux expected over long ngEHT baselines. Performance for 2017 is taken over 2\,GHz of bandwidth and the observed median sensitivity of ALMA and LMT during EHT April 2017 observations, and this is extended to the full projected bandwidth at 230\,GHz for ALMA-II and LMT-II. NOEMA is calculated for a 12-element array under nominal weather conditions,
and the small ngEHT remote site is evaluated for moderately poor line-of-sight opacity of 0.6.}}\label{fig:anchor_sites2}
\end{SCfigure*}

Current EHT images are already exceptionally rich scientifically. Following system engineering practices, and referencing the STM in figure \ref{tab:stm}, we propose to extend the scientific potential of ground-based mm VLBI observations by quadrupling the current recorded instantaneous bandwidth of the EHT, adding a 345 GHz capability, and incorporating new sites to the existing array. This last possibility stems from the important realization that single large apertures in the array (phased ALMA in Chile, the Large Millimeter Telescope in Mexico, and future phased NOEMA in France) provide such high sensitivity that adding small-diameter dishes in ideal geographic locations can dramatically improve imaging fidelity -- even for sites where the atmospheric conditions are more variable than is typical for current mm/submm facilities (\autoref{fig:anchor_sites2}).
By roughly doubling the number of antennas in the EHT 
through addition of several new small diameter dishes as well as new stations
 ngEHT can reconstruct not just images of extraordinary detail, but movies of the dynamics near the black hole event horizon. 

\begin{figure}[h]
    \centering
    \includegraphics[width=1.00 \textwidth]{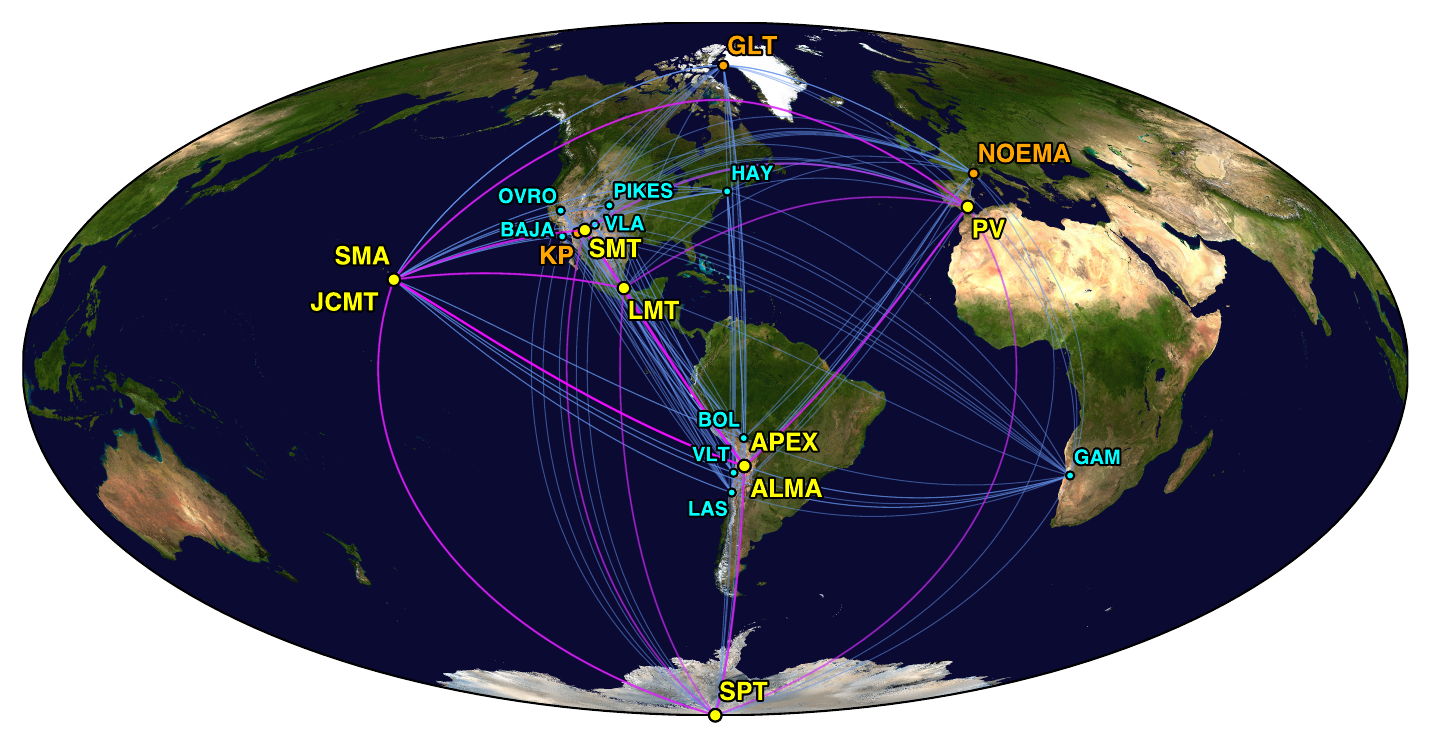}
    \caption{Distribution of stations around the globe.  Stations that participated in the EHT2017 observing campaign are labeled in yellow, while the additional stations that will be present in the EHT2020 array are labeled in orange. Several possible new site locations for the ngEHT are labeled in cyan.
    Current EHT2017 baselines are shown in magenta.
    }
    \label{fig:globe}
\end{figure}

\begin{figure}[h]
    \centering
    \includegraphics[width=0.49\textwidth]{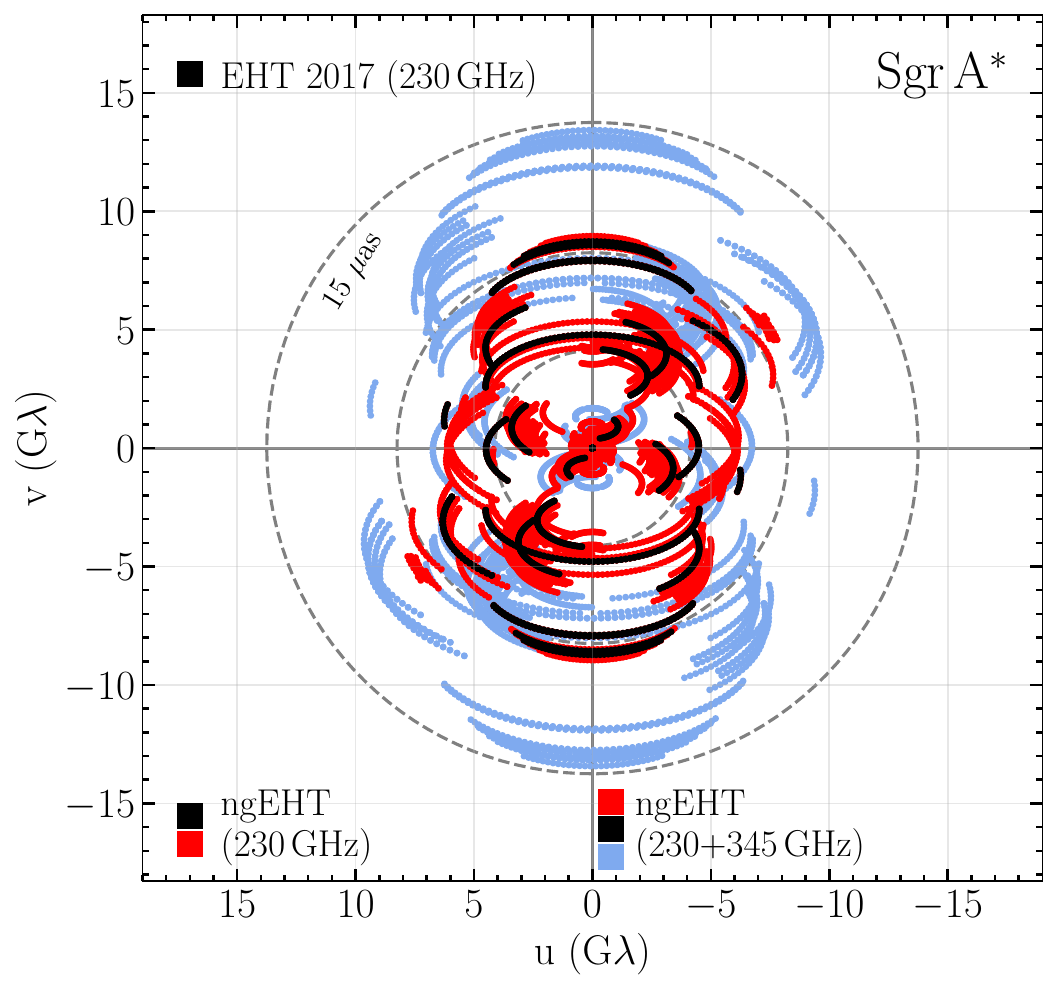}
    \includegraphics[width=0.49\textwidth]{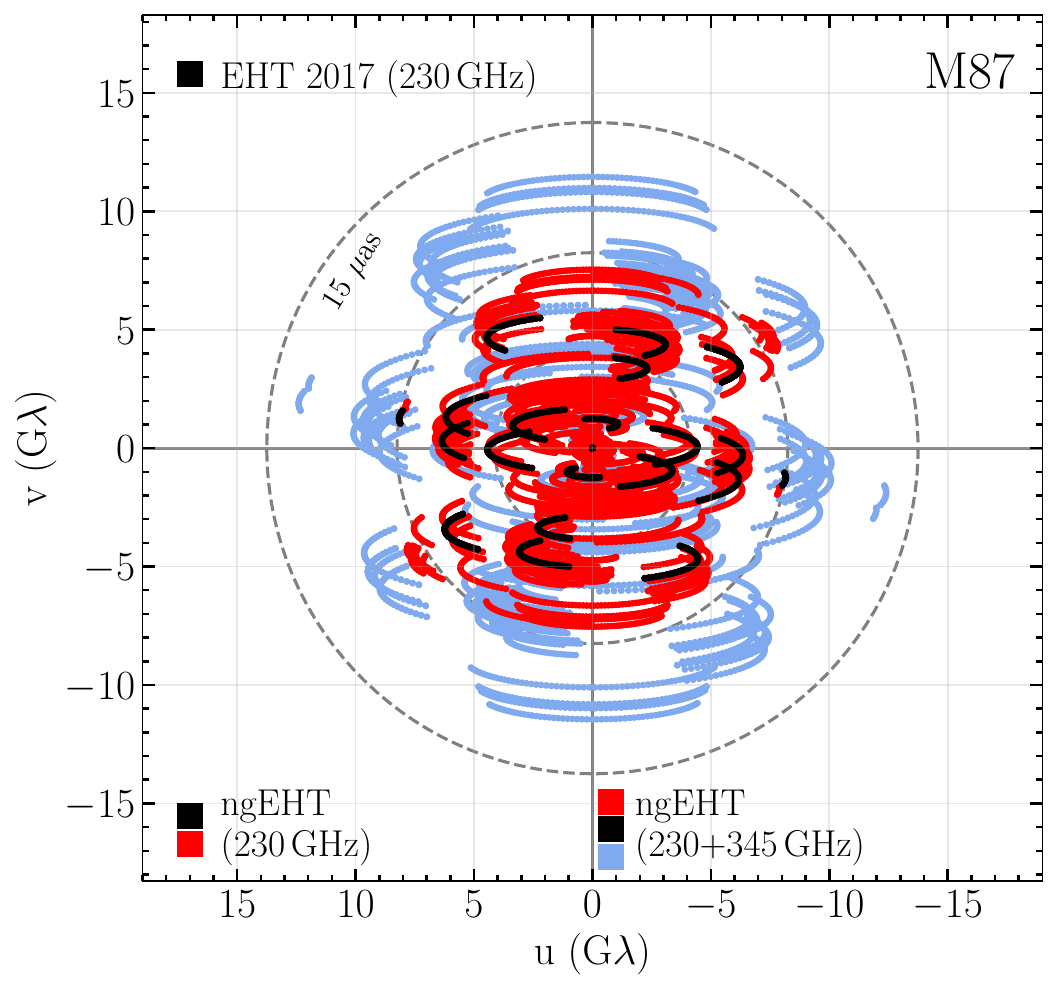}
    \caption{Fourier space coverage of the EHT primary  science sources for the EHT2017 array \citep{paperii} and for the proposed expanded ngEHT array as shown in Figure \ref{fig:globe}. OVRO, HAY, KP and GAM sites are excluded from 345\,GHz operations.}
    \label{fig:uv_coverage}
\end{figure}

\subsection{New sites and dishes}

At the bandwidth projected for the ngEHT, antenna diameters between 6 and 12~m will be suitably sensitive for new nodes in the array. The Greenland Telescope is an example of a successful relocation of a 12~m ALMA dish, and a similar dish is being relocated to Kitt Peak in Arizona. In ngEHT Phase~I, designs for new dishes will be explored using approaches that have been successfully used for sub-mm class antennas for the SMA and \mbox{ALMA~\citep{Moran1984, Lazareff2001}}.  

Candidate locations for newly designed dishes will be selected based on weather for sub-mm observing, existing infrastructure, and improvement to the spatial frequency coverage of the array \citep{Palumbo_2019,Lal_2010}.
 Small dishes are particularly effective close to major ngEHT anchor sites. An example of an expanded ngEHT array, feasible by 2027, is shown in \autoref{fig:globe}.
Corresponding improvements in the \uv-coverage for the EHT science targets are shown in \autoref{fig:uv_coverage}.

\subsection{Receiver and VLBI back end}
The EHT presently samples 4\,GHz bandwidth in dual polarization and two sidebands for a total of 16\,GHz. This corresponds  to 64\,Gbps for two-bit recording and Nyquist sampling.  This matches ALMA's current bandwidth, though efforts are underway to double the ALMA bandwidth in each sideband over the next decade.   The majority of the other EHT sites already employ receivers with 8\,GHz sidebands, and those that do not are typically in the process of upgrading.  A doubling of bandwidth per sideband for the EHT would result in a record rate of 128\,Gbps.

The ability to simultaneously observe the 1.3\,mm (230\,GHz) and 0.87\,mm (345\,GHz) EHT observing bands dramatically improves imaging and movie rendering capability of the EHT (Figures \ref{fig:sgra_movie}, \ref{fig:ngEHT_jet}) as well as polarization observations of, for example, Faraday rotation. The ngEHT with 8~GHz per sideband, dual polarization, and simultaneous dual band 230/345\,GHz capability requires a recording rate of 256\,Gbps. A dual-band EHT receiver that will serve as the prototype for other suitable telescopes in the EHT array will be first installed and commissioned on the LMT, to immediately enhances the high resolution capability of the EHT.
1.3\,mm and 0.87\,mm wavelength sideband-separating dual-polarization mixers have been built for facilities such as ALMA and the Institut de Radioastronomie Millimetrique (IRAM). A dual-frequency receiver will be deployed to additional telescopes in the array.

New back end development takes advantage of advances in FPGA and ADC capabilities. We are developing a back end capable of processing four 8-GHz bands using 16\,Gsps samplers. This is a total of 128 Gbps at 2-bit quantization, available in a compact rack mount box. The European DBBC3 is another platform capable of supporting 128 Gbps using 8\,Gsps samplers.
A key development is to match the back end rate in data capture and transport.
The current Mark 6 VLBI recorder \citep{Whitney_2013} has in the lab operated at close to twice its design target of 16 Gbps, due largely to the steady increase in hard disk density and throughput. One path is to further develop the Mark 6 for reliable operation at higher speeds in the field. COTS recording solutions that use Field Programmable Gate Arrays (FPGA) for high-speed parallel data flow are emerging and will also be explored.

\subsection{Data transport and processing}

The current EHT and VLBI practice of recording data and physically transporting them to a central location for correlation has severe disadvantages. This includes inability to verify in real-time that an experiment has been setup correctly and is working.  Data storage disks used by the EHT are large, expensive and take a long time to back up, which means there is a very real risk of data loss in shipment.  Observations from the South Pole Station, absolutely key for \sgra, are hobbled because shipment of disks recorded in April takes effectively six months to return to the correlation center. This causes time delays in the data analysis and requires that the data from all stations be saved for cross-correlation at one time.

With the potential eight-fold increase in collected data volume over the current systems, the EHT will develop an innovative way to store and consolidate data for processing. Free space laser communication presents an alternative. Data rates of many Terabits/sec are possible. It is an extremely attractive technology to consider for an entirely new paradigm of EHT operations, supporting real time data transport and correlation.  The TBIRD system, developed at MIT Lincoln Laboratory \citep{RobinsonB.S2018TIDT}, is one example that will be investigated and evaluated.

\subsection{Algorithmic developments, array optimization}

Development of new analysis methods, specially designed to address challenges in EHT data, were critical for producing the first images of a black hole. Due to the extreme sparsity and calibration uncertainties for high frequency VLBI, recovering an image of \m87 required the development of new methods that simultaneously performed imaging and calibration, as well as the design of techniques to validate the result \citep{paperiv}. For the ngEHT to meet its Science Goals successfully, new techniques for the reconstruction of movies of \sgra \mbox{\citep{Johnson_2017,Bouman_2018}}, polarimetric images of \m87 \citep{Chael_2016,Akiyama_2017}, and fringe detection and calibration strategies tailored for a hybrid array with globally distributed anchor stations and several small dishes \citep{Blackburn2019} are required. Ultra-wide bandwidths and simultaneous observing at 230 and 345 GHz will also motivate the development of new techniques for multi-frequency imaging, scattering mitigation toward \sgra \citep{Johnson_2016}, and atmospheric phase transfer~\citep{Honma2003,Reid2014}. The analysis methods will be coupled with realistic array data simulation~\citep{Blecher2017} incorporating dynamical source models, instrument characteristics, as well as weather and scheduling models in order to optimize ngEHT array design through the use of performance metrics. For example at 230 GHz, using the \uv filling fraction metric \citep{Palumbo_2019} at 500 $\mu$as field-of-view with 100 $\mu$as resolution, the EHT2017 samples 20\% of \sgra and 25\% of \m87, while the ngEHT samples 71\% of \sgra and 99\% of \m87 (see \autoref{fig:uv_coverage}).

\section{Organization - Array \& Partners}

The EHT Collaboration (EHTC), coordinates and conducts EHT observation campaigns and sets the agenda for EHT science and development.  A Memorandum of Collaboration binds a group of thirteen stakeholders that currently include (Academia Sinica Institute of Astronomy and Astrophysics, University of Arizona, University of Chicago, East Asian Observatory, Goethe-Universitaet Frankfurt, Institut de Radioastronomie Millimetrique, Large Millimeter Telescope Alfonso Serrano, Max Planck Institute for Radioastronomy, MIT Haystack Observatory, National Astronomical Observatory of Japan, Perimeter Institute for Theoretical Physics, Radboud University, and the Smithsonian Astrophysical Observatory).  Development of future directions and design is a collaborative and ongoing process within the EHTC, and this APC white paper describes activities that require support of US involvement in this timely effort.  The current EHTC international agreement has also enabled, through partnership with ALMA, an {\it open-skies} policy for the general astronomy community to use the EHT array for high resolution, high sensitivity applications.  It is anticipated that the current EHT agreement will be continued, possibly evolving to accommodate a larger operational component with more observing epochs and a doubling of antennas in the array.  Retaining open-skies access to the EHT array is similarly anticipated with commensurate impact on the global astronomy infrastructure.

Over the course of EHT build-out, two previously used sites (Caltech Submillimeter Observatory and the Combined Array for Research in Millimeter-wave Astronomy) were decommissioned.  This was balanced by the inclusion of new facilities, and the proposed deployment of a number of new modest-diameter dishes will minimize impact on the ngEHT to the loss of existing facilities over the coming decade should that occur.

\section{Schedule and cost}

We envisage the design and eventual implementation of this proposed ngEHT as two separate phases, both of which exist within the overall organization of the EHT project. Phase~I will be a design process that optimizes a systematic approach to defining science goals and development specifications, leading to Preliminary Design Reviews (PDR) and Critical Design Reviews (CDR) for main elements by $\sim$2023. A critical component of Phase~I will be the selection and preparation of new sites. Phase~II will be a build out and array augmentation with new VLBI systems, with in-place equipment by $\sim$2027, and commissioning and observations thereafter. Increased interim capability through the staged introduction of sites, increased bandwidth, and the build-out of e.g. 345 GHz capabilities at EHT stations will provide new scientific opportunities throughout Phase~I and Phase~II, as was the case for the initial EHT over the previous decade.

Assuming a site has basic infrastructure and roads suitable for a crane, a new EHT node will cost between \$2.5M for a small, refurbished antenna or as much as \$10M for a new 12-m antenna.  The costs for a timing reference, recording equipment, a receiver and the peripheral RF and test equipment are common to both cases and total about \$1.3M.  The costs for a shelter and solar power plant will be between \$0.8M and \$1.0M depending on the local weather.  One full-time equivalent engineering salary is factored into the totals. From these numbers we project a requirement of $\sim$\$10M for upgrades to existing sites and  $\sim$\$40--\$100M for the addition of 8 new sites for the completed construction of the ngEHT array, placing it at the intersection of medium ($>$\$20M) and large ($>$\$70M) ground project categories. Operational cost is anticipated at the level of $\sim$\$5M annually. As with the initial EHT, the next generation EHT would be comprised of a global network of radio telescopes with observations and scientific utilization managed through worldwide collaboration, and is expected to be funded through multiple international sources and in kind contributions from partners and ngEHT stakeholder institutions.  Based on current EHT support, we estimate the net cost to US funding agencies will be $\sim$1/2 of the total projected ngEHT cost.  This will depend on future arrangements with international EHT partners, however given the world-wide impact of the first EHT results, we anticipate continued strong interest and engagement from the international community.

\section{Summary}
To build upon the success of the EHT in imaging the SMBH at the center of M87 on horizon scales, we propose the design and implementation of a ground-based Next Generation EHT (ngEHT). This instrument will double the number of antennas in the array, incorporate a dual-frequency capability, more than double the sensitivity, and increase the dynamic range by more than one order of magnitude over the existing EHT. This will enable fundamental questions to be tackled, both in physics (e.g., the space-time metric around a rotating black hole and deviations from the predictions of GR) and in astrophysics (e.g. the launching mechanism of jets in AGNs and the role of magnetic fields in black hole accretion). 

The EHT effort has delivered the first black hole image.  The ngEHT will fulfill the promise of a newly emerging field of research in astronomy and physics: precision imaging and time resolution of black holes on horizon scales.

\pagebreak

\printbibliography

@inproceedings{RobinsonB.S2018TIDT,
pages = {105240V--105240V-6},
volume = {10524},
publisher = {SPIE},
isbn = {9781510615335},
year = {2018},
title = {TeraByte InfraRed Delivery (TBIRD): a demonstration of large-volume direct-to-Earth data transfer from low-Earth orbit},
language = {eng},
author = {Robinson, B. S and Boroson, D. M and Schieler, C. M and Khatri, F. I and Guldner, O and Constantine, S and Shih, T and Burnside, J. W and Bilyeu, B. C and Hakimi, F and Garg, A and Allen, G and Clements, E and Cornwell, D. M},
}

@ARTICLE{Johannsen_2010,
   author = {{Johannsen}, T. and {Psaltis}, D.},
    title = "{Testing the No-hair Theorem with Observations in the Electromagnetic Spectrum. II. Black Hole Images}",
  journal = {\apj},
archivePrefix = "arXiv",
   eprint = {1005.1931},
     year = 2010,
    month = jul,
   volume = 718,
    pages = {446-454},
      doi = {10.1088/0004-637X/718/1/446},
   adsurl = {http://adsabs.harvard.edu/abs/2010ApJ...718..446J}
}

@ARTICLE{Tchekhovskoy_2011,
   author = {{Tchekhovskoy}, A. and {Narayan}, R. and {McKinney}, J.~C.},
    title = "{Efficient generation of jets from magnetically arrested accretion on a rapidly spinning black hole}",
  journal = {\mnras},
archivePrefix = "arXiv",
   eprint = {1108.0412},
     year = 2011,
    month = nov,
   volume = 418,
    pages = {L79-L83},
      doi = {10.1111/j.1745-3933.2011.01147.x},
   adsurl = {http://adsabs.harvard.edu/abs/2011MNRAS.418L..79T}
}

@ARTICLE{Palumbo_2019,
       author = {{Palumbo}, D. C.~M. and {Doeleman}, S. S. and
         {Johnson}, M. D. and {Bouman}, K. L. and {Chael}, A. A.},
        title = "{Metrics and Motivations for Earth-Space VLBI: Time-resolving Sgr A* with the Event Horizon Telescope}",
      journal = {\apj},
         year = "2019",
        month = "Aug",
       volume = {881},
       number = {1},
          eid = {62},
        pages = {62},
          doi = {10.3847/1538-4357/ab2bed},
archivePrefix = {arXiv},
       eprint = {1906.08828},
       adsurl = {https://ui.adsabs.harvard.edu/abs/2019ApJ...881...62P}
}

@article{Blandford:1982vy,
   author = {{Blandford}, R.~D. and {Payne}, D.~G.},
    title = "{Hydromagnetic flows from accretion discs and the production of radio jets}",
  journal = {\mnras},
     year = 1982,
    month = jun,
   volume = 199,
    pages = {883-903},
      doi = {10.1093/mnras/199.4.883},
   adsurl = {https://ui.adsabs.harvard.edu/abs/1982MNRAS.199..883B}
}

@ARTICLE{Lal_2010,
       author = {{Lal}, D. V. and {Lobanov}, A. P. and
         {Jim{\'e}nez-Monferrer}, S.},
        title = "{Array configuration studies for the Square Kilometre Array - Implementation of figures of merit based on spatial dynamic range}",
      journal = {arXiv e-prints},
     keywords = {Astrophysics - Instrumentation and Methods for Astrophysics},
         year = "2010",
        month = "Jan",
%          eid = {arXiv:1001.1477},
%        pages = {arXiv:1001.1477},
archivePrefix = {arXiv},
       eprint = {1001.1477},
       adsurl = {https://ui.adsabs.harvard.edu/abs/2010arXiv1001.1477L},
      adsnote = {Provided by the SAO/NASA Astrophysics Data System}
}

@ARTICLE{Balbus_1998,
   author = {{Balbus}, S.~A. and {Hawley}, J.~F.},
    title = "{Instability, turbulence, and enhanced transport in accretion disks}",
  journal = {Reviews of Modern Physics},
     year = 1998,
    month = jan,
   volume = 70,
    pages = {1-53},
      doi = {10.1103/RevModPhys.70.1},
   adsurl = {http://adsabs.harvard.edu/abs/1998RvMP...70....1B}
}

@techreport{Lazareff2001,
author = {Lazareff, B.},
booktitle = {ALMA Memo 395},
pages = {1--14},
title = {{Alignment Tolerances for ALMA Optics}},
year = {2001}
}

@techreport{Moran1984,
author = {Moran, J.M. and Elvis, M.S. and Fazio, G.G. and Ho, P.T.P. and Myers, P.C. and Reid, M.J. and Willner, S.P.},
booktitle = {SMA Memo 0},
pages = {1--145},
title = {{A Submillimeter-Wavelength Telescope Array: Scientific, Technical, and Strategic Issues}},
year = {1984}
}

@ARTICLE{Chael_2019,
       author = {{Chael}, A. A. and {Narayan}, R. and {Johnson}, M. D.},
        title = "{Two-temperature, Magnetically Arrested Disc simulations of the jet from the supermassive black hole in M87}",
      journal = {\mnras},
         year = "2019",
        month = "Jun",
       volume = {486},
       number = {2},
        pages = {2873-2895},
          doi = {10.1093/mnras/stz988},
archivePrefix = {arXiv},
       eprint = {1810.01983},
       adsurl = {https://ui.adsabs.harvard.edu/abs/2019MNRAS.486.2873C}
}

@ARTICLE{Blandford_1977,
   author = {{Blandford}, R.~D. and {Znajek}, R.~L.},
    title = "{Electromagnetic extraction of energy from Kerr black holes}",
  journal = {\mnras},
     year = 1977,
    month = may,
   volume = 179,
    pages = {433-456},
      doi = {10.1093/mnras/179.3.433},
   adsurl = {http://adsabs.harvard.edu/abs/1977MNRAS.179..433B}
}

@ARTICLE{Broderick_2006,
   author = {{Broderick}, A.~E. and {Loeb}, A.},
    title = "{Imaging optically-thin hotspots near the black hole horizon of Sgr A* at radio and near-infrared wavelengths}",
  journal = {\mnras},
   eprint = {astro-ph/0509237},
     year = 2006,
    month = apr,
   volume = 367,
    pages = {905-916},
      doi = {10.1111/j.1365-2966.2006.10152.x},
   adsurl = {http://adsabs.harvard.edu/abs/2006MNRAS.367..905B}
}

@ARTICLE{Doeleman_2008,
   author = {{Doeleman}, S.~S. and {Weintroub}, J. and {Rogers}, A.~E.~E. and 
	{Plambeck}, R. and {Freund}, R. and {Tilanus}, R.~P.~J. and 
	{Friberg}, P. and {Ziurys}, L.~M. and {Moran}, J.~M. and {Corey}, B. and 
	{Young}, K.~H. and {Smythe}, D.~L. and {Titus}, M. and {Marrone}, D.~P. and 
	{Cappallo}, R.~J. and {Bock}, D.~C.-J. and {Bower}, G.~C. and 
	{Chamberlin}, R. and {Davis}, G.~R. and {Krichbaum}, T.~P. and 
	{Lamb}, J. and {Maness}, H. and {Niell}, A.~E. and {Roy}, A. and 
	{Strittmatter}, P. and {Werthimer}, D. and {Whitney}, A.~R. and 
	{Woody}, D.},
    title = "{Event-horizon-scale structure in the supermassive black hole candidate at the Galactic Centre}",
  journal = {\nat},
archivePrefix = "arXiv",
   eprint = {0809.2442},
     year = 2008,
    month = sep,
   volume = 455,
    pages = {78-80},
      doi = {10.1038/nature07245},
   adsurl = {http://adsabs.harvard.edu/abs/2008Natur.455...78D}
}

@ARTICLE{Dodds-Eden_2009,
   author = {{Dodds-Eden}, K. and {Porquet}, D. and {Trap}, G. and {Quataert}, E. and 
	{Haubois}, X. and {Gillessen}, S. and {Grosso}, N. and {Pantin}, E. and 
	{Falcke}, H. and {Rouan}, D. and {Genzel}, R. and {Hasinger}, G. and 
	{Goldwurm}, A. and {Yusef-Zadeh}, F. and {Clenet}, Y. and {Trippe}, S. and 
	{Lagage}, P.-O. and {Bartko}, H. and {Eisenhauer}, F. and {Ott}, T. and 
	{Paumard}, T. and {Perrin}, G. and {Yuan}, F. and {Fritz}, T.~K. and 
	{Mascetti}, L.},
    title = "{Evidence for X-Ray Synchrotron Emission from Simultaneous Mid-Infrared to X-Ray Observations of a Strong Sgr A* Flare}",
  journal = {\apj},
archivePrefix = "arXiv",
   eprint = {0903.3416},
     year = 2009,
    month = jun,
   volume = 698,
    pages = {676-692},
      doi = {10.1088/0004-637X/698/1/676},
   adsurl = {http://adsabs.harvard.edu/abs/2009ApJ...698..676D}
}

@ARTICLE{Doeleman_2009,
   author = {{Doeleman}, S.~S. and {Fish}, V.~L. and {Broderick}, A.~E. and 
	{Loeb}, A. and {Rogers}, A.~E.~E.},
    title = "{Detecting Flaring Structures in Sagittarius A* with High-Frequency VLBI}",
  journal = {\apj},
archivePrefix = "arXiv",
   eprint = {0809.3424},
     year = 2009,
    month = apr,
   volume = 695,
    pages = {59-74},
      doi = {10.1088/0004-637X/695/1/59},
   adsurl = {http://adsabs.harvard.edu/abs/2009ApJ...695...59D}
}

@ARTICLE{Doeleman_2012,
   author = {{Doeleman}, S.~S. and {Fish}, V.~L. and {Schenck}, D.~E. and 
	{Beaudoin}, C. and {Blundell}, R. and {Bower}, G.~C. and {Broderick}, A.~E. and 
	{Chamberlin}, R. and {Freund}, R. and {Friberg}, P. and {Gurwell}, M.~A. and 
	{Ho}, P.~T.~P. and {Honma}, M. and {Inoue}, M. and {Krichbaum}, T.~P. and 
	{Lamb}, J. and {Loeb}, A. and {Lonsdale}, C. and {Marrone}, D.~P. and 
	{Moran}, J.~M. and {Oyama}, T. and {Plambeck}, R. and {Primiani}, R.~A. and 
	{Rogers}, A.~E.~E. and {Smythe}, D.~L. and {SooHoo}, J. and 
	{Strittmatter}, P. and {Tilanus}, R.~P.~J. and {Titus}, M. and 
	{Weintroub}, J. and {Wright}, M. and {Young}, K.~H. and {Ziurys}, L.~M.
	},
    title = "{Jet-Launching Structure Resolved Near the Supermassive Black Hole in M87}",
  journal = {Science},
archivePrefix = "arXiv",
   eprint = {1210.6132},
     year = 2012,
    month = oct,
   volume = 338,
    pages = {355},
      doi = {10.1126/science.1224768},
   adsurl = {http://adsabs.harvard.edu/abs/2012Sci...338..355D}
}

@ARTICLE{Whitney_2013,
   author = {{Whitney}, A.~R. and {Beaudoin}, C.~J. and {Cappallo}, R.~J. and 
	{Corey}, B.~E. and {Crew}, G.~B. and {Doeleman}, S.~S. and {Lapsley}, D.~E. and 
	{Hinton}, A.~A. and {McWhirter}, S.~R. and {Niell}, A.~E. and 
	{Rogers}, A.~E.~E. and {Ruszczyk}, C.~A. and {Smythe}, D.~L. and 
	{SooHoo}, J. and {Titus}, M.~A.},
    title = "{Demonstration of a 16 Gbps Station$^{-1}$ Broadband-RF VLBI System}",
  journal = {\pasp},
     year = 2013,
    month = feb,
   volume = 125,
    pages = {196},
      doi = {10.1086/669718},
   adsurl = {http://adsabs.harvard.edu/abs/2013PASP..125..196W}
}

@ARTICLE{Broderick_2014,
   author = {{Broderick}, A.~E. and {Johannsen}, T. and {Loeb}, A. and {Psaltis}, D.
	},
    title = "{Testing the No-hair Theorem with Event Horizon Telescope Observations of Sagittarius A*}",
  journal = {\apj},
archivePrefix = "arXiv",
   eprint = {1311.5564},
     year = 2014,
    month = mar,
   volume = 784,
      eid = {7},
    pages = {7},
      doi = {10.1088/0004-637X/784/1/7},
   adsurl = {http://cdsads.u-strasbg.fr/abs/2014ApJ...784....7B}
}

@ARTICLE{Chael_2016,
   author = {{Chael}, A.~A. and {Johnson}, M.~D. and {Narayan}, R. and {Doeleman}, S.~S. and 
	{Wardle}, J.~F.~C. and {Bouman}, K.~L.},
    title = "{High-resolution Linear Polarimetric Imaging for the Event Horizon Telescope}",
  journal = {\apj},
archivePrefix = "arXiv",
   eprint = {1605.06156},
     year = 2016,
    month = sep,
   volume = 829,
      eid = {11},
    pages = {11},
      doi = {10.3847/0004-637X/829/1/11},
   adsurl = {http://adsabs.harvard.edu/abs/2016ApJ...829...11C}
}

@ARTICLE{Johnson_2016,
   author = {{Johnson}, M.~D.},
    title = "{Stochastic Optics: A Scattering Mitigation Framework for Radio Interferometric Imaging}",
  journal = {\apj},
archivePrefix = "arXiv",
   eprint = {1610.05326},
     year = 2016,
    month = dec,
   volume = 833,
      eid = {74},
    pages = {74},
      doi = {10.3847/1538-4357/833/1/74},
   adsurl = {http://adsabs.harvard.edu/abs/2016ApJ...833...74J}
}

@ARTICLE{Akiyama_2017,
   author = {{Akiyama}, K. and {Ikeda}, S. and {Pleau}, M. and {Fish}, V.~L. and 
	{Tazaki}, F. and {Kuramochi}, K. and {Broderick}, A.~E. and 
	{Dexter}, J. and {Mo{\'s}cibrodzka}, M. and {Gowanlock}, M. and 
	{Honma}, M. and {Doeleman}, S.~S.},
    title = "{Superresolution Full-polarimetric Imaging for Radio Interferometry with Sparse Modeling}",
  journal = {\aj},
archivePrefix = "arXiv",
   eprint = {1702.00424},
     year = 2017,
    month = apr,
   volume = 153,
      eid = {159},
    pages = {159},
      doi = {10.3847/1538-3881/aa6302},
   adsurl = {http://adsabs.harvard.edu/abs/2017AJ....153..159A}
}

@Article{Boccardi_2017,
author="Boccardi, B.
and Krichbaum, T. P.
and Ros, E.
and Zensus, J. A.",
title="Radio observations of active galactic nuclei with mm-VLBI",
journal="The Astronomy and Astrophysics Review",
year="2017",
month="Nov",
day="06",
volume="25",
number="1",
pages="4",
issn="1432-0754",
doi="10.1007/s00159-017-0105-6",
% url="https://doi.org/10.1007/s00159-017-0105-6"
}

@ARTICLE{Johnson_2017,
   author = {{Johnson}, M.~D. and {Bouman}, K.~L. and {Blackburn}, L. and 
	{Chael}, A.~A. and {Rosen}, J. and {Shiokawa}, H. and {Roelofs}, F. and 
	{Akiyama}, K. and {Fish}, V.~L. and {Doeleman}, S.~S.},
    title = "{Dynamical Imaging with Interferometry}",
  journal = {\apj},
archivePrefix = "arXiv",
   eprint = {1711.01286},
     year = 2017,
    month = dec,
   volume = 850,
      eid = {172},
    pages = {172},
      doi = {10.3847/1538-4357/aa97dd},
   adsurl = {http://cdsads.u-strasbg.fr/abs/2017ApJ...850..172J}
}

@ARTICLE{Bouman_2018, 
author={{Bouman}, K. and {Johnson}, M. and {Dalca}, A. and {Chael}, A. and {Roelofs}, R. and {Doeleman}. S. and {Freeman}, W.~T.}, 
journal={IEEE Transactions on Computational Imaging}, 
title={Reconstructing Video of Time-Varying Sources from Radio Interferometric Measurements}, 
year={2018}, 
volume={4}, 
number={4}, 
pages={512-527}, 
doi={10.1109/TCI.2018.2838452}, 
ISSN={2333-9403}, 
month={},}

@ARTICLE{gravity_redshift_2018,
       author = {{Gravity Collaboration} and {Abuter}, R. and {Amorim}, A. and
         {Anugu}, N. and {Baub{\"o}ck}, M. and {Benisty}, M. and
         {Berger}, J.~P. and {Blind}, N. and {Bonnet}, H. and {Brandner}, W. and
         {Buron}, A. and {Collin}, C. and {Chapron}, F. and {Cl{\'e}net}, Y. and
         {Coud{\'e} Du Foresto}, V. and {de Zeeuw}, P.~T. and {Deen}, C. and
         {Delplancke-Str{\"o}bele}, F. and {Dembet}, R. and {Dexter}, J. and
         {Duvert}, G. and {Eckart}, A. and {Eisenhauer}, F. and {Finger}, G. and
         {F{\"o}rster Schreiber}, N.~M. and {F{\'e}dou}, P. and {Garcia}, P. and
         {Garcia Lopez}, R. and {Gao}, F. and {Gendron}, E. and {Genzel}, R. and
         {Gillessen}, S. and {Gordo}, P. and {Habibi}, M. and {Haubois}, X. and
         {Haug}, M. and {Hau{\ss}mann}, F. and {Henning}, Th. and {Hippler}, S. and
         {Horrobin}, M. and {Hubert}, Z. and {Hubin}, N. and
         {Jimenez Rosales}, A. and {Jochum}, L. and {Jocou}, K. and
         {Kaufer}, A. and {Kellner}, S. and {Kendrew}, S. and {Kervella}, P. and
         {Kok}, Y. and {Kulas}, M. and {Lacour}, S. and {Lapeyr{\`e}re}, V. and
         {Lazareff}, B. and {Le Bouquin}, J. -B. and {L{\'e}na}, P. and
         {Lippa}, M. and {Lenzen}, R. and {M{\'e}rand}, A. and {M{\"u}ler}, E. and
         {Neumann}, U. and {Ott}, T. and {Palanca}, L. and {Paumard}, T. and
         {Pasquini}, L. and {Perraut}, K. and {Perrin}, G. and {Pfuhl}, O. and
         {Plewa}, P.~M. and {Rabien}, S. and {Ram{\'\i}rez}, A. and {Ramos}, J. and
         {Rau}, C. and {Rodr{\'\i}guez-Coira}, G. and {Rohloff}, R. -R. and
         {Rousset}, G. and {Sanchez-Bermudez}, J. and {Scheithauer}, S. and
         {Sch{\"o}ller}, M. and {Schuler}, N. and {Spyromilio}, J. and
         {Straub}, O. and {Straubmeier}, C. and {Sturm}, E. and
         {Tacconi}, L.~J. and {Tristram}, K.~R.~W. and {Vincent}, F. and
         {von Fellenberg}, S. and {Wank}, I. and {Waisberg}, I. and
         {Widmann}, F. and {Wieprecht}, E. and {Wiest}, M. and {Wiezorrek}, E. and
         {Woillez}, J. and {Yazici}, S. and {Ziegler}, D. and {Zins}, G.},
        title = "{Detection of the gravitational redshift in the orbit of the star S2 near the Galactic centre massive black hole}",
      journal = {\aap},
         year = "2018",
        month = "Jul",
       volume = {615},
          eid = {L15},
        pages = {L15},
          doi = {10.1051/0004-6361/201833718},
archivePrefix = {arXiv},
       eprint = {1807.09409},
       adsurl = {https://ui.adsabs.harvard.edu/\#abs/2018A&A...615L..15G}
}

@ARTICLE{Gravity_2018,
   author = {{Gravity Collaboration} and {Abuter}, R. and {Amorim}, A. and 
	{Baub{\"o}ck}, M. and {Berger}, J.~P. and {Bonnet}, H. and {Brandner}, W. and 
	{Cl{\'e}net}, Y. and {Coud{\'e} Du Foresto}, V. and {de Zeeuw}, P.~T. and 
	{Deen}, C. and {Dexter}, J. and {Duvert}, G. and {Eckart}, A. and 
	{Eisenhauer}, F. and {F{\"o}rster Schreiber}, N.~M. and {Garcia}, P. and 
	{Gao}, F. and {Gendron}, E. and {Genzel}, R. and {Gillessen}, S. and 
	{Guajardo}, P. and {Habibi}, M. and {Haubois}, X. and {Henning}, T. and 
	{Hippler}, S. and {Horrobin}, M. and {Huber}, A. and {Jim{\'e}nez-Rosales}, A. and 
	{Jocou}, L. and {Kervella}, P. and {Lacour}, S. and {Lapeyr{\`e}re}, V. and 
	{Lazareff}, B. and {Le Bouquin}, J.-B. and {L{\'e}na}, P. and 
	{Lippa}, M. and {Ott}, T. and {Panduro}, J. and {Paumard}, T. and 
	{Perraut}, K. and {Perrin}, G. and {Pfuhl}, O. and {Plewa}, P.~M. and 
	{Rabien}, S. and {Rodr{\'{\i}}guez-Coira}, G. and {Rousset}, G. and 
	{Sternberg}, A. and {Straub}, O. and {Straubmeier}, C. and {Sturm}, E. and 
	{Tacconi}, L.~J. and {Vincent}, F. and {von Fellenberg}, S. and 
	{Waisberg}, I. and {Widmann}, F. and {Wieprecht}, E. and {Wiezorrek}, E. and 
	{Woillez}, J. and {Yazici}, S.},
    title = "{Detection of orbital motions near the last stable circular orbit of the massive black hole SgrA*}",
  journal = {\aap},
archivePrefix = "arXiv",
   eprint = {1810.12641},
     year = 2018,
    month = oct,
   volume = 618,
      eid = {L10},
    pages = {L10},
      doi = {10.1051/0004-6361/201834294},
   adsurl = {http://adsabs.harvard.edu/abs/2018A\%26A...618L..10G}
}

@ARTICLE{paperi,
   author = {{Event Horizon Telescope Collaboration} and {Akiyama}, K. and 
	{Alberdi}, A. and {Alef}, W. and {Asada}, K. and {Azulay}, R. and 
	{Baczko}, A.-K. and {Ball}, D. and {Balokovi{\'c}}, M. and {Barrett}, J. and et al.},
    title = "{First M87 Event Horizon Telescope Results. I. The Shadow of the Supermassive Black Hole}",
  journal = {\apjl},
     year = 2019,
    month = apr,
   volume = 875,
      eid = {L1},
    pages = {L1},
      doi = {10.3847/2041-8213/ab0ec7},
   adsurl = {http://adsabs.harvard.edu/abs/2019ApJ...875L...1E}
}

@ARTICLE{paperii,
   author = {{Event Horizon Telescope Collaboration} and {Akiyama}, K. and 
	{Alberdi}, A. and {Alef}, W. and {Asada}, K. and {Azulay}, R. and 
	{Baczko}, A.-K. and {Ball}, D. and {Balokovi{\'c}}, M. and {Barrett}, J. and et al.},
    title = "{First M87 Event Horizon Telescope Results. II. Array and Instrumentation}",
  journal = {\apjl},
     year = 2019,
    month = apr,
   volume = 875,
      eid = {L2},
    pages = {L2},
      doi = {10.3847/2041-8213/ab0c96},
   adsurl = {http://adsabs.harvard.edu/abs/2019ApJ...875L...2E}
}

@ARTICLE{paperiv,
   author = {{Event Horizon Telescope Collaboration} and {Akiyama}, K. and 
	{Alberdi}, A. and {Alef}, W. and {Asada}, K. and {Azulay}, R. and 
	{Baczko}, A.-K. and {Ball}, D. and {Balokovi{\'c}}, M. and {Barrett}, J. and et al.},
    title = "{First M87 Event Horizon Telescope Results. IV. Imaging the Central Supermassive Black Hole}",
  journal = {\apjl},
     year = 2019,
    month = apr,
   volume = 875,
      eid = {L4},
    pages = {L4},
      doi = {10.3847/2041-8213/ab0e85},
   adsurl = {https://ui.adsabs.harvard.edu/abs/2019ApJ...875L...4E}
}

@ARTICLE{Psaltis_2015,
   author = {{Psaltis}, D. and {{\"O}zel}, F. and {Chan}, C.-K. and {Marrone}, D.~P.
	},
    title = "{A General Relativistic Null Hypothesis Test with Event Horizon Telescope Observations of the Black Hole Shadow in Sgr A*}",
  journal = {\apj},
archivePrefix = "arXiv",
   eprint = {1411.1454},
     year = 2015,
    month = dec,
   volume = 814,
      eid = {115},
    pages = {115},
      doi = {10.1088/0004-637X/814/2/115},
   adsurl = {http://adsabs.harvard.edu/abs/2015ApJ...814..115P}
}

@ARTICLE{Hawking_1972,
   author = {{Hawking}, S.~W.},
    title = "{Black holes in general relativity}",
  journal = {Communications in Mathematical Physics},
     year = 1972,
    month = jun,
   volume = 25,
    pages = {152-166},
      doi = {10.1007/BF01877517},
   adsurl = {http://adsabs.harvard.edu/abs/1972CMaPh..25..152H}
}

@ARTICLE{Blackburn2019,
% 	doi = {10.3847/1538-4357/ab328d},
% %	url = {https://doi.org/10.3847%2F1538-4357%2Fab328d},
% 	year = 2019,
% 	month = {aug},
% 	publisher = {American Astronomical Society},
% 	volume = {882},
% 	number = {1},
% 	pages = {23},
%       author = {{Blackburn}, L and {Chan}, C.-K. and {Crew}, G. B. and
%          {Fish}, V. L. and {Issaoun}, S. and {Johnson}, M. D. and
%          {Wielgus}, M. and {Akiyama}, K. and {Barrett}, J. and
%          {Bouman}, K. L. and {Cappallo}, R. and {Chael}, A. A. and
%          {Janssen}, M. and {Lonsdale}, C. J. and {Doeleman}, S. S.},
% 	title = {{EHT}-{HOPS} Pipeline for Millimeter {VLBI} Data Reduction},
% 	archivePrefix = {arXiv},
%       eprint = {1903.08832},
% 	journal = {\apj},
%       adsurl = {https://ui.adsabs.harvard.edu/abs/2019arXiv190308832B},
%       adsnote = {Provided by the SAO/NASA Astrophysics Data System}
% }

@ARTICLE{Honma2003,
       author = {{Honma}, M. and {Fujii}, T. and {Hirota}, T. and
         {Horiai}, K. and {Iwadate}, K. and {Jike}, T. and
         {Kameya}, O. and {Kamohara}, R. and {Kan-Ya}, Y. and
         {Kawaguchi}, N.},
        title = "{First Fringe Detection with VERA's Dual-Beam System and Its Phase-Referencing Capability}",
      journal = {\pasj},
         year = "2003",
        month = "Jun",
       volume = {55},
        pages = {L57-L60},
          doi = {10.1093/pasj/55.4.L57},
archivePrefix = {arXiv},
       eprint = {astro-ph/0306060},
       adsurl = {https://ui.adsabs.harvard.edu/abs/2003PASJ...55L..57H}
}

@ARTICLE{Reid2014,
       author = {{Reid}, M.~J. and {Honma}, M.},
        title = "{Microarcsecond Radio Astrometry}",
      journal = {\araa},
         year = "2014",
        month = "Aug",
       volume = {52},
        pages = {339-372},
          doi = {10.1146/annurev-astro-081913-040006},
archivePrefix = {arXiv},
       eprint = {1312.2871},
       adsurl = {https://ui.adsabs.harvard.edu/abs/2014ARA&A..52..339R}
}

@ARTICLE{Blecher2017,
       author = {{Blecher}, T. and {Deane}, R. and {Bernardi}, G. and
         {Smirnov}, O.},
        title = "{MEQSILHOUETTE: a mm-VLBI observation and signal corruption simulator}",
      journal = {\mnras},
         year = "2017",
        month = "Jan",
       volume = {464},
       number = {1},
        pages = {143-151},
          doi = {10.1093/mnras/stw2311},
archivePrefix = {arXiv},
       eprint = {1608.04521},
       adsurl = {https://ui.adsabs.harvard.edu/abs/2017MNRAS.464..143B}
}

\end{document}